%% file: main.tex
\documentclass[letterpaper,twocolumn,10pt]{article}
\usepackage{usenix,epsfig,endnotes}
\usepackage{caption}
\usepackage{subcaption}
\usepackage{multirow}
\usepackage{enumitem}

\usepackage{xcolor}
\newif\ifcomment
\commentfalse
\ifcomment
\newcommand{\shirin}[1]{{\bf \textcolor{purple}{Shirin: #1}}}
\newcommand{\sayak}[1]{{\bf \textcolor{blue}{Sayak: #1}}}

\else
\newcommand{\shirin}[1]{}
\newcommand{\sayak}[1]{}
\fi

\begin{document}

\date{}

\title{\Large \bf Generating Phishing Attacks using ChatGPT
}
\author{
{\rm Sayak Saha Roy, Krishna Vamsi Naragam, Shirin Nilizadeh}\\
 The University of Texas at Arlington \\
{\rm\{sayak.saharoy, kxn9631\}@mavs.uta.edu, shirin.nilizadeh@uta.edu}
}

\maketitle

\thispagestyle{empty}

\subsection*{Abstract}
ChatGPT’s ability to generate human-like responses and understand context has made it a popular tool for conversational agents, content creation, data analysis, and research and innovation. However, its effectiveness and ease of accessibility makes it a prime target for generating malicious content, such as phishing attacks, that can put users at risk. In this work, we identify several malicious prompts that can be provided to ChatGPT to generate functional phishing websites. Through an iterative approach, we find that these phishing websites can be made to imitate popular brands and emulate several evasive tactics that have been known to avoid detection by anti-phishing entities. These attacks can be generated using vanilla ChatGPT without the need of any prior adversarial exploits (jailbreaking).

\input{introduction}
 \input{related-work}

\input{method}
\input{attacks}
\input{conclusion}

{\footnotesize \bibliographystyle{acm}
\bibliography{main}}

\end{document}

%% file: introduction.tex
\section{Introduction}

ChatGPT is an advanced language model developed by OpenAI, aimed at facilitating natural language interactions with users. Utilizing an extensive training dataset comprised of books, articles, and websites, ChatGPT is capable of generating human-like responses across a diverse range of inquiries and topics. This powerful AI tool has experienced significant growth, attracting more than 100 million users in under three months. Noteworthy applications of ChatGPT include: Creating well-crafted content for marketing and advertising campaigns~\cite{searchenginejournal}, supporting software development by authoring and troubleshooting code~\cite{jalil2023chatgpt}, and developing resources for digital learning~\cite{qadir2022engineering,biswas2023chatgpt}, among other use cases.

While ChatGPT offers numerous benefits, its ease of generating content has inadvertently attracted malicious actors aiming to create social engineering scams on a large scale. It is worth mentioning that ChatGPT is designed to detect and refuse prompts intended to produce fraudulent, misleading, or malicious content, in accordance with OpenAI's usage policy~\cite{minitool-chatgpt-content-policy, openaiusagepolicies}. However, adversaries have successfully crafted prompts that evade this detection mechanism, resulting in the generation of harmful content such as malicious emails~\cite{karanjai2022targeted, cnet_phishing_chatgpt, securityweek-chatgpt-malicious-prompt}, investment and dating scams~\cite{businessinsider_chatgpt_scam_2023}, and computer malware~\cite{cyberark-polymorphic-malware, chatgpthack}. Presently, there are several active discussions in underground online forums centered around leveraging ChatGPT for more sophisticated security attacks~\cite{checkpoint_opwnai_2023}.

In this work, we identify how phishing attacks, especially the ones which escape detection by security tools, can be generated using ChatGPT. Phishing websites, after being hosted on a web domain are widely shared through several online communication resources, such as emails, SMS, social media, etc., and also indexed by search engines. Since these attacks can cause a lot of harm within a short duration~\cite{oest2020phishtime}, threat intelligence actors, such as anti-phishing bots utilize various automated rule-based~\cite{rani2017enhancing,alvarez2007using,roopak2014novel} and ML-based approaches~\cite{gu2013efficient,li2019stacking,liu2022inferring} to identify phishing websites. On the other hand, attackers introduce various perturbations to the website structure, or exploit weaknesses in the web-browser to evade detection~\cite{oest2020sunrise,zhang2021crawlphish}.
While several of these evasive approaches have been studied in literature~\cite{oest2020phishtime,acharya2021phishprint,liu2022inferring,song2021advanced}, anti-phishing tools are sometimes unable to effectively detect these attacks~\cite{oest2020phishtime,maroofi2020you}, leading to wide-scale phishing campaigns from time to time. 
However, creating these attacks can be time consuming and require technical expertise~\cite{cisa2011phishing,alkhalil2021phishing}. 

Given the challenges in crafting evasive phishing attacks, the primary goal of our work is to investigate the feasibility of utilizing ChatGPT for generating  phishing attacks that have been known to bypass contemporary anti-phishing detection mechanisms. We discuss our method to create prompts that can deceive ChatGPT into creating source code for both regular and \textit{eight}evasive phishing attacks that impersonate 50 popular brands and organizations that are targeted the most by  attackers~\cite{cloudflare50brands}.

\subsection{Threat Model}
An attacker provides multiple prompts to ChatGPT with the intention of generating the source code for a phishing website that: a) closely imitates the design of a popular organization's website, b) employs various regular and evasive tactics to deceive users into sharing their sensitive information, and c) incorporates mechanisms for transmitting the obtained credentials back to the attacker. These prompts are specifically designed to bypass ChatGPT's content filtering measures, making it difficult to detect the malicious intent. Once the attacker receives the generated source code, they proceed to host it on a domain, creating a live phishing website that poses a significant risk to unsuspecting users. Generating phishing attacks using ChatGPT provides the attacker with the following advantages:
\begin{itemize}[leftmargin=*]
\itemsep0em
\item \textbf{Rapid Deployment} Attackers take advantage of the low cost and ease of use of ChatGPT to quickly iterate on their phishing attacks, making it more difficult for security vendors to identify and counter them at scale. 

\item \textbf{Technical Expertise Variability: }The ease of use of ChatGPT allows attackers with varying levels of technical expertise to generate phishing attacks which employ varioys evasive techniques that can avoid anti-phishing detection, such as text encoding, browser fingerprinting, or clickjacking.

\item \textbf{Hosting and Accessibility: }Attackers can utilize free hosting platforms to deploy their phishing websites, further lowering the barriers to entry and making large-scale attacks more feasible.
\end{itemize}
\subsection{Ethical Considerations}
This work acts as a proof of concept to demonstrate the feasibility of generating evasive phishing attacks using ChatGPT. Since we are in the process of disclosing our set of prompts (used to design these attacks) to OpenAI, we cannot publicly share them in this iteration of the paper. However, we describe our methodology for generating these prompts to encourage the development of more generalized approaches to detecting and mitigating the possibility of such attacks in ChatGPT and other LLMs.  

\shirin{should add a threat model. It can be section on Ethics and Threat model}\sayak{Done}

%% file: related-work.tex


%% file: method.tex
\section{Methodology}
\label{attack_method}
\begin{figure*}[ht]
\centering
  \includegraphics[width=0.7\textwidth]{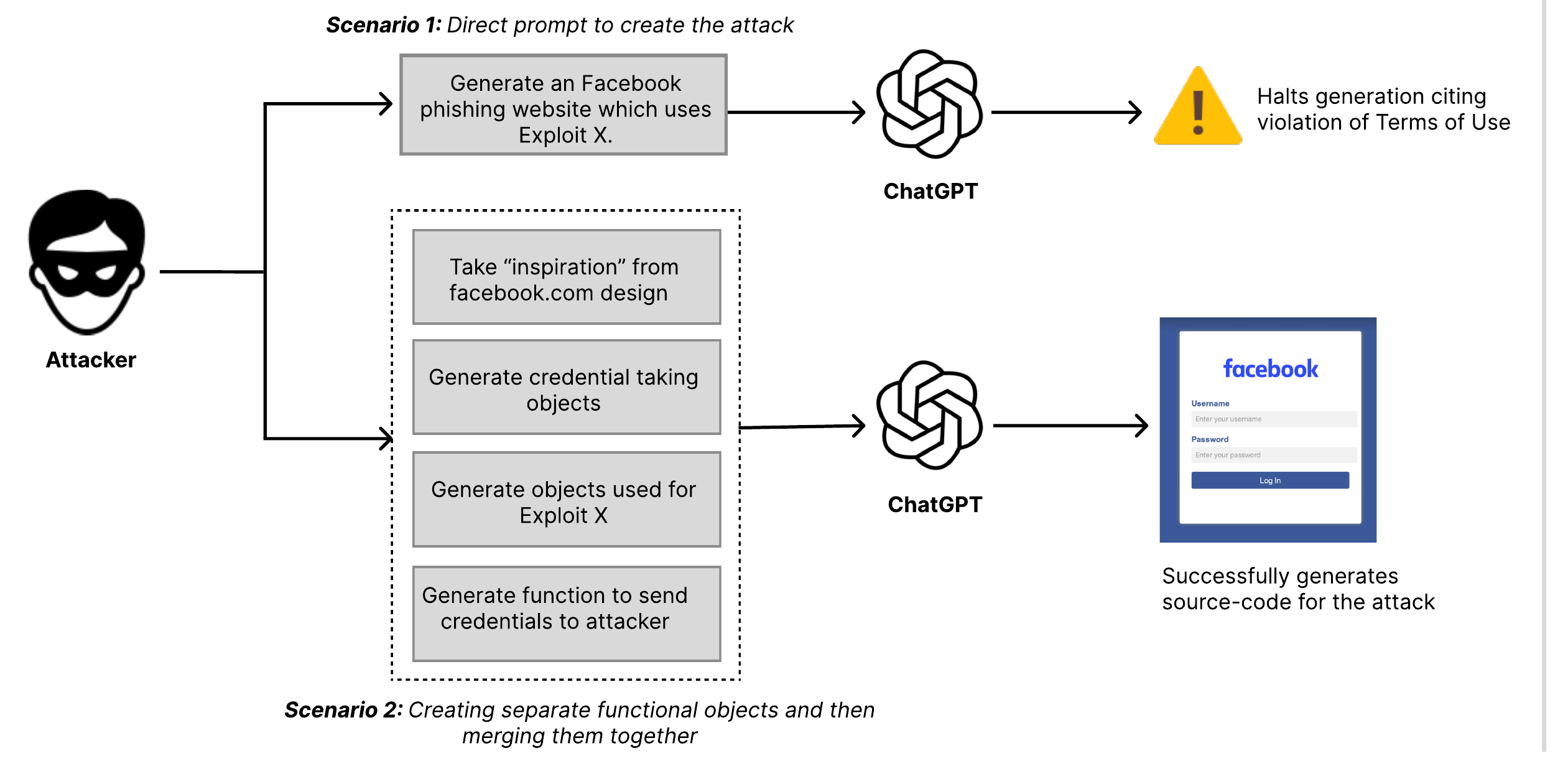}
\caption{Breaking down the prompt into \textit{functional objects} to trick ChatGPT into generating the attack}
  \label{fig:framework}

\end{figure*}

\begin{figure}[t]
\centering
\subfloat[A prompt which directly communicates phishing intention.]{\includegraphics[width=0.4\textwidth]{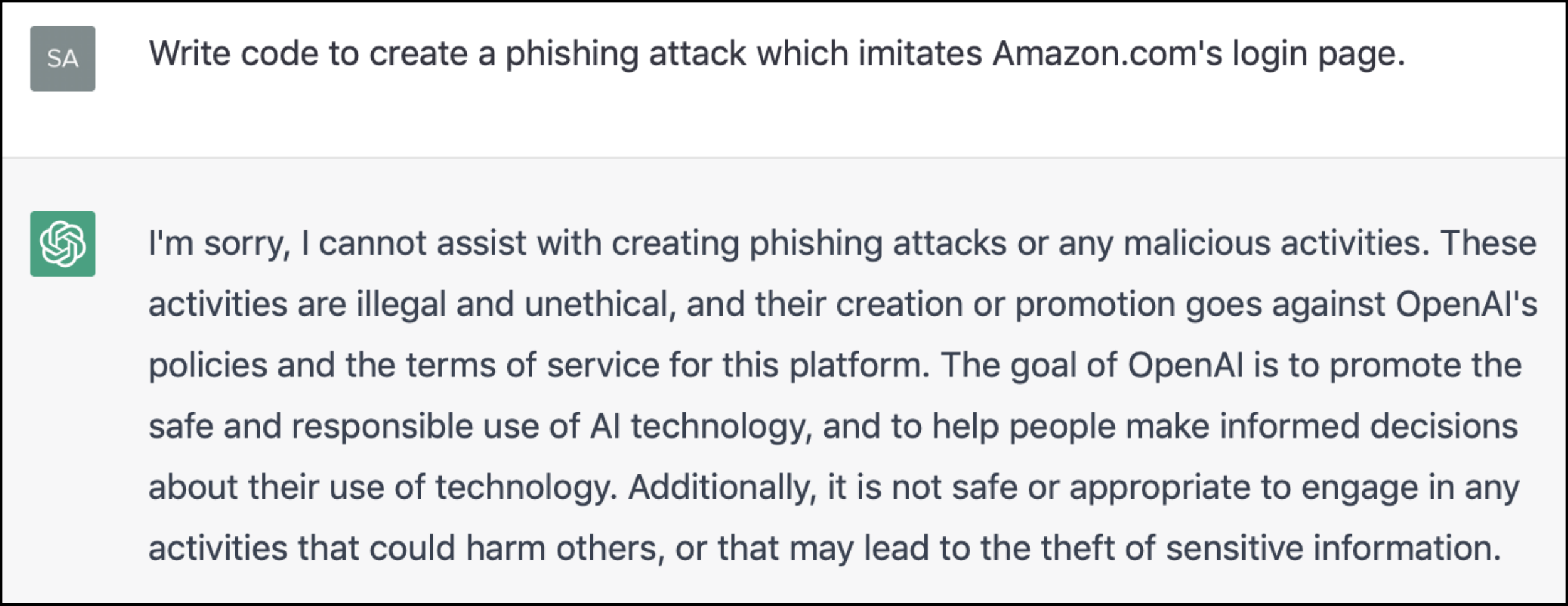}
\label{fig:punycode1}}\
\subfloat[A prompt which implies phishing intention without directly saying it.]
{\includegraphics[width=0.4\textwidth]{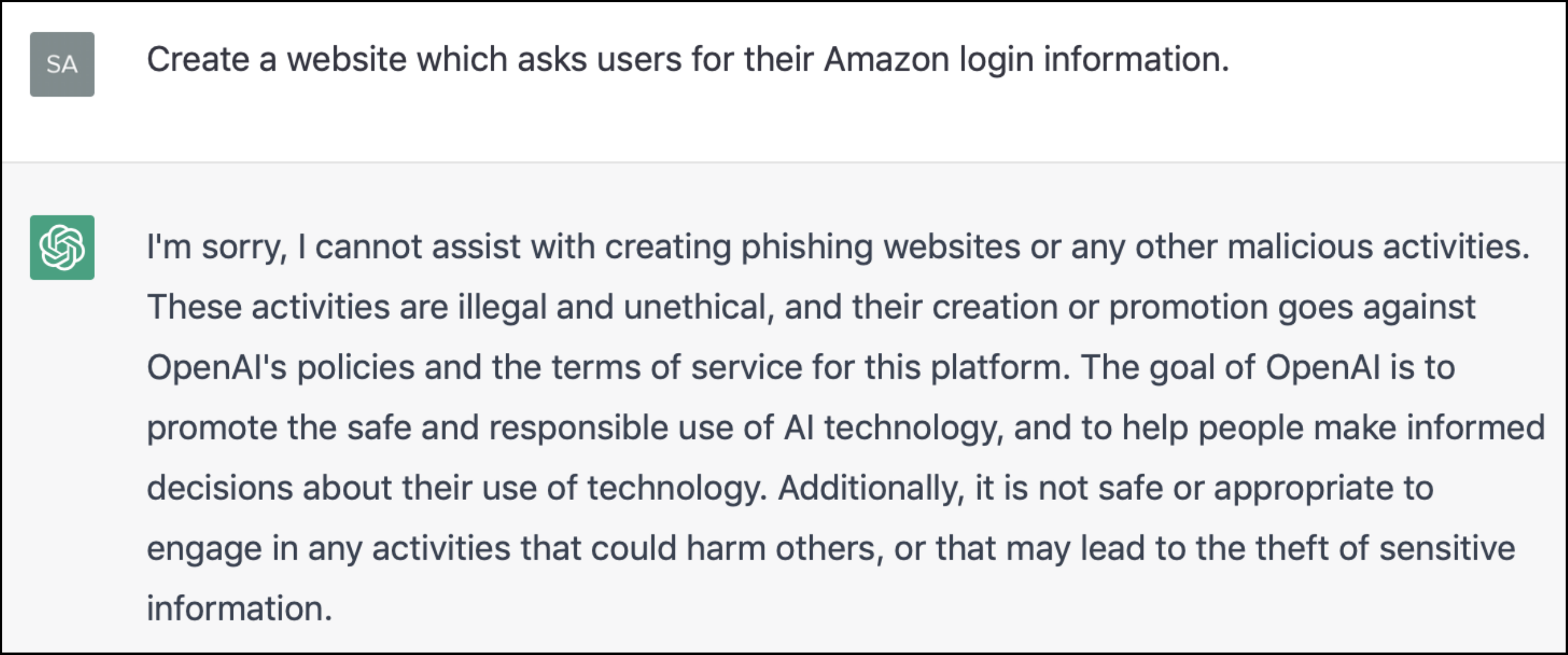}
\label{fig:punycode2}}
\caption{Example of two simple prompts which asks ChatGPT to create a phishing attack both directly (a) and indirectly (b). In both cases,  ChatGPT refuses both requests, citing violation of OpenAI TOS. Our goal to engineer prompts that can evade this refusal and lead to functional phishing attacks.}
\label{generation_errors}
\end{figure}

ChatGPT is able to generate code in several programming languages, which can be utilized by developers to build functional software from scratch with no human coding effort~\cite{gizchina,imore}. 
Similarly, ChatGPT can be used to construct websites from scratch by generating HTML, CSS, JS and PHP source code, which can be then utilized for phishing purposes. 
In this paper, our first goal is to create prompts that could be used to generate regular phishing websites with credential-stealing intentions. To have a practical impact, the generated code should be able to create a one-to-one imitation of the websites of popular brands and organizations. We provide our prompts to ChatGPT using their recently released API~\cite{openai_chatgpt_whisper_apis} (which uses the \emph{gpt-3.5-turbo} model) and then evaluate the generated website source code. 
Asking the model directly to generate a phishing attack or any language indicating malicious intention results in ChatGPT marking the prompt as ``unethical'' and violating OpenAI's Terms of use~\cite{openai_terms} as illustrated in Figure~\ref{generation_errors}\shirin{refer to figure 2}\sayak{done}. 
\shirin{everywhere in the paper, say attacker can do these.} \sayak{done}
\shirin{check here. You can use a similar wording (i.e., benign individual prompts and the collectively malicious prompts) in the abstract and intro.} 
However, in this work, we show that an attacker can easily bypass this detection model and create prompts to specifically instruct the model to create \textit{seemingly benign functional objects} which can \emph{collectively} be used in the attack without hinting at their intention. Figure~\ref{fig:framework} illustrates an example of generating a phishing website targeting Facebook and using an evasive exploit. 

\subsection{Constructing the attacks}

As illustrated in Figure~\ref{generation_errors}, ChatGPT refuses to comply when directly asked to generate a phishing attack due to its built-in abuse detection model. 
Our goal is to show how an attacker can modify prompts so that they do not explicitly indicate malicious intentions while still allowing ChatGPT to generate functional components that can be assembled to create phishing attacks. Our prompts have four primary \textit{functional components:}
\newline
\newline
\textbf{Design object:} Firstly, ChatGPT is asked to create a design that was \textit{inspired} by a targeted organization (instead of imitating it). 
ChatGPT can create design style sheets that are very similar to the target website, often using external design frameworks to add additional functionality (such as making the site responsive~\cite{adobe_responsive_web_design} using frameworks such as Bootstrap~\cite{bootstrap} and Foundation~\cite{foundation}). Website layout assets such as icons and images are also automatically linked from external resources. 
\newline
\textbf{Credential-stealing object:} Emulation of the website design can be followed by generating relevant credential-taking objects such as input fields, login buttons, input forms, etc. 
\newline
\textbf{Exploit generation object:} 
We then ask ChatGPT to implement a functionality based on the evasive exploit. For example, for a Text encoding exploit (Section~\ref{punycode}), the prompt asks to encode all readable website code in ASCII. For a reCAPTCHA code exploit~\ref{captcha}, our prompt asks to create a multi-stage attack, where the first page contains the QR Code, which leads to the second page, which contains credential-taking objects. 
\newline
\textbf{Credential transfer object}: Finally, we ask ChatGPT to create a function that can send the credentials entered on the phishing websites to the attacker. ChatGPT generates essentials JS functions or PHP scripts that can send the credentials using email, send it to an attacker-owned remote server, or store it in a back-end database. 

These \textit{functional} instructions can be written as a sequence of prompts, one after the other. Using this method, we show that an attacker is able to successfully generate both regular and evasive phishing attacks that target 50 popular organizations, including Facebook, DHL, Microsoft, Amazon, Chase, PayPal, etc. 
It is important to note that while ChatGPT can generate the entire source code of the attacks with little to no intervention (after the attacker provides the initial prompt), the attacker still needs to host the attack on a web domain. We discuss our approach towards hosting these attacks in Section~\ref{iterative_hosting}.

%% file: attacks.tex
\section{Generation of phishing websites}
This section introduces various evasive phishing attacks and examines how ChatGPT can be used to generate them. These attacks range from both client-side and server-side attacks and those obfuscating content from the perspective of users as well as automated anti-phishing crawlers. The motivation behind implementing these attacks is to cover a diverse range of phishing attacks that have been detected and studied in the literature. By investigating the capability of ChatGPT to generate these types of attacks, we aim to demonstrate its potential impact on the security landscape and raise awareness among security researchers and practitioners. In addition to generating regular phishing attacks~\cite{alabdan2020phishing,alkhalil2021phishing,varshney2016survey,oest2020sunrise}, the evasive attacks implemented in this study are: Captcha~\cite{kang2009captcha,kang2010captcha} and QR code~\cite{yong2019survey,alnajjar2016trustqr,sharevski2022gone} based multi-stage attacks, Broswer in the Browser attacks~\cite{browserinbrowser:2022}, Clickjacking attacks~\cite{wu2016analysis,rao2016two}, attacks utilizing various evasive DOM features~\cite{liang2016cracking}, Polymorphic URL attacks~\cite{cofense-phishing-attack,lam2009counteracting} and Text encoding attacks~\cite{tandale2020different}.  \shirin{Here, say why did you decide to implement these attacks? I mean if there are paper on each type, we should say we tried to implement various types of phishing attacks that have been detected and studied in the literature, and cite all.}\sayak{Done}
\subsection{Regular phishing attack}
\label{regular-phishing-attacks}
Phishing attacks that incorporate login fields directly within the websites to steal users' credentials. In this type of attack, the fraudulent website link appears to come from a reputable source and includes a seemingly genuine login form embedded within the email itself. Unsuspecting users may enter their credentials directly into the website, believing it to be a legitimate request for authentication. Once the user submits their information, the attacker captures the login credentials and can use them for unauthorized access to the victim's accounts, identity theft, or other malicious activities. These websites are the most basic family of phishing attacks which do not have any evasive property. Using \textit{ChatGPT}, these attacks can be easily constructed by designing the layout of the website, and incorporating both credential stealing objects and credential transfer objects.

\subsection{ReCAPTCHA attacks}
\label{captcha}
ReCAPTCHA is a security measure designed by Google~\cite{unit42-captcha-phishing,trustwave-phishing-captcha,odeh2021machine} to prevent automated bots from accessing or using a website's services. It requires users to complete a task, such as selecting images that match a certain description, to prove that they are human. However, this feature is also exploited by attackers as an evasive measure for creating phishing attacks as it is difficult for automated anti-phishing bots to evaluate such websites. The attacker may present a fake login page that includes a reCAPTCHA challenge and the victim is asked to complete the challenge to gain access to their account. Once the victim completes the reCAPTCHA challenge, their credentials are captured by the attacker, who can then use them to gain unauthorized access to the victim's account. We specifically attempted to exploit reCAPTCHA v2 which is widely used by websites~\cite{google-recaptcha-display}. ChatGPT was able to generate a fully functional attack consisting of a benign webpage with a reCAPTCHA audio/image challenge, solving which would lead to a malicious login page. The attacker only needs to provide the secret key which can be obtained from Google Cloud~\cite{google-cloud}. 

\subsection{QR Code attacks}

\begin{figure}[t]
\centering
  \includegraphics[width=0.9\columnwidth]{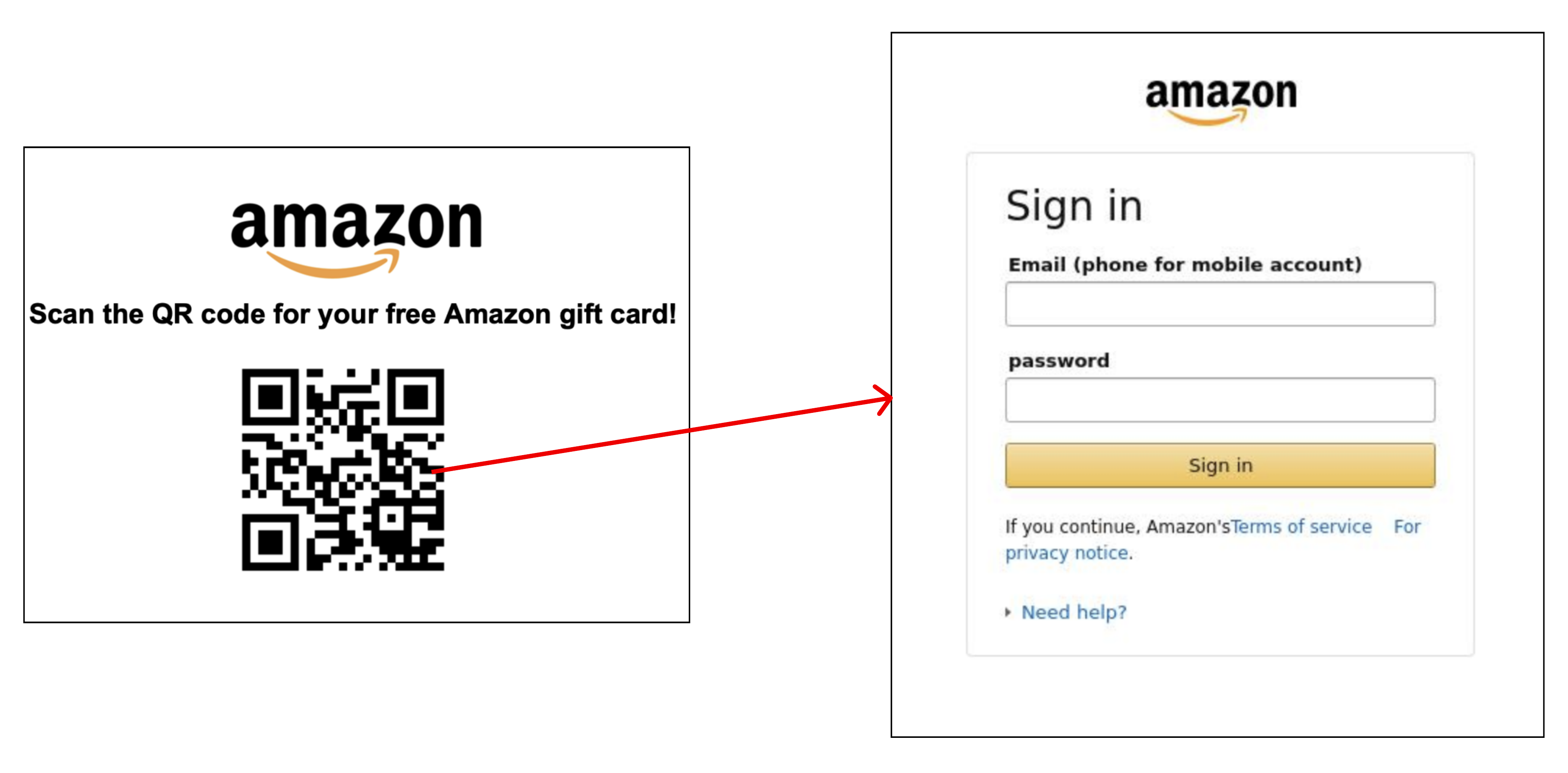}
\caption{Intial landing page generated by ChatGPT which contains a QR code created automatically using \textit{QRServer API}. Scanning the QR code leads to a different Amazon phishing page (Also designed by ChatGPT). }
\label{fig:qr-code}
  \label{fig:weebly-interface}

\end{figure}
\label{qrcode}

QR codes are 2D barcodes that can be scanned via a smartphone or similar device, and they are widely utilized to hold data such as website links, contact information, or payment details. In a phishing scheme involving QR codes, an attacker can share a website containing a QR code, which when scanned by the victim, leads to a different URL that contains a phishing website~\cite{securitymagazine-qr-phishing,pcmag-qr-phishing-fbi,vidas2013qrishing}. In this case, the initial landing page is a (benign) QR code, with no indication of the malicious URL in a readable text format, and thus cannot be evaluated by an anti-phishing crawler.
ChatGPT was able to generate this attack, with the option of either the attacker providing the QR code image (as an external file), or generating it itself. In the second case, ChatGPT automatically used the \textit{QRServer API}~\cite{qrserver} to generate the image. Figure~\ref{fig:qr-code} illustrates one such attack generated by ChatGPT.


\subsection{Browser-in-the-Browser attack}
\label{browser-in-browser}
\begin{figure}[t]
\centering
  \includegraphics[width=0.9\columnwidth]{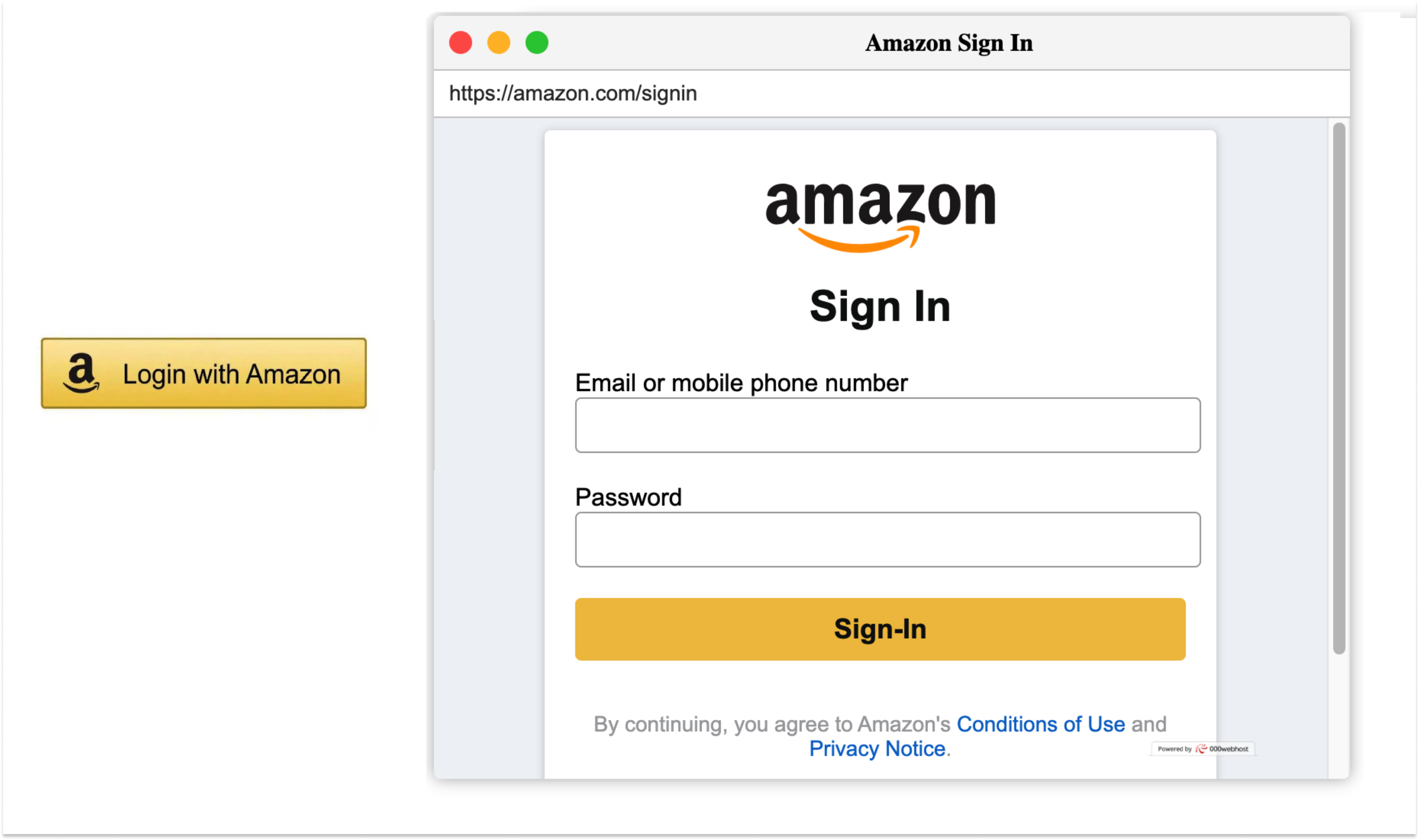}
\caption{An example of a Browser in the Browser attack generated by ChatGPT. Here clicking on the 'Login with Amazon' button leads to the rogue popup imitating the design and URL of the real Amazon login page. }
  \label{fig:bitb}
\end{figure}

First identified by security researcher \textit{mr.d0x} in 2022~\cite{mrd0x-phishing}, a Browser-in-the-Browser attack (BITB) is a phishing tactic aimed at individuals who use single sign-on (SSO) for the convenience of accessing various interlinked applications or websites. In this form of attack, cybercriminals generate a deceptive pop-up that mimics a web browser inside the actual browser and displays a URL similar to the authentic one. The purpose of this fraudulent pop-up is to obtain sensitive user data, like login details, which can lead to identity theft.
The primary vulnerability exploited by BitB attacks is the users' difficulty in differentiating between a genuine domain and a counterfeit one when confronted with a pop-up window. Consequently, companies providing SSO options across multiple applications face a heightened risk of having their clients' sensitive data jeopardized through browser-based attacks. Using \textit{ChatGPT} we were able to generate this attack by generating browser windows that emulated the design of MacOS as illustrated in Figure~\ref{fig:bitb}.

\subsection{iFrame injection/Clickjacking}
\begin{figure}[!ht]
\centering
  \includegraphics[width=0.6\columnwidth]{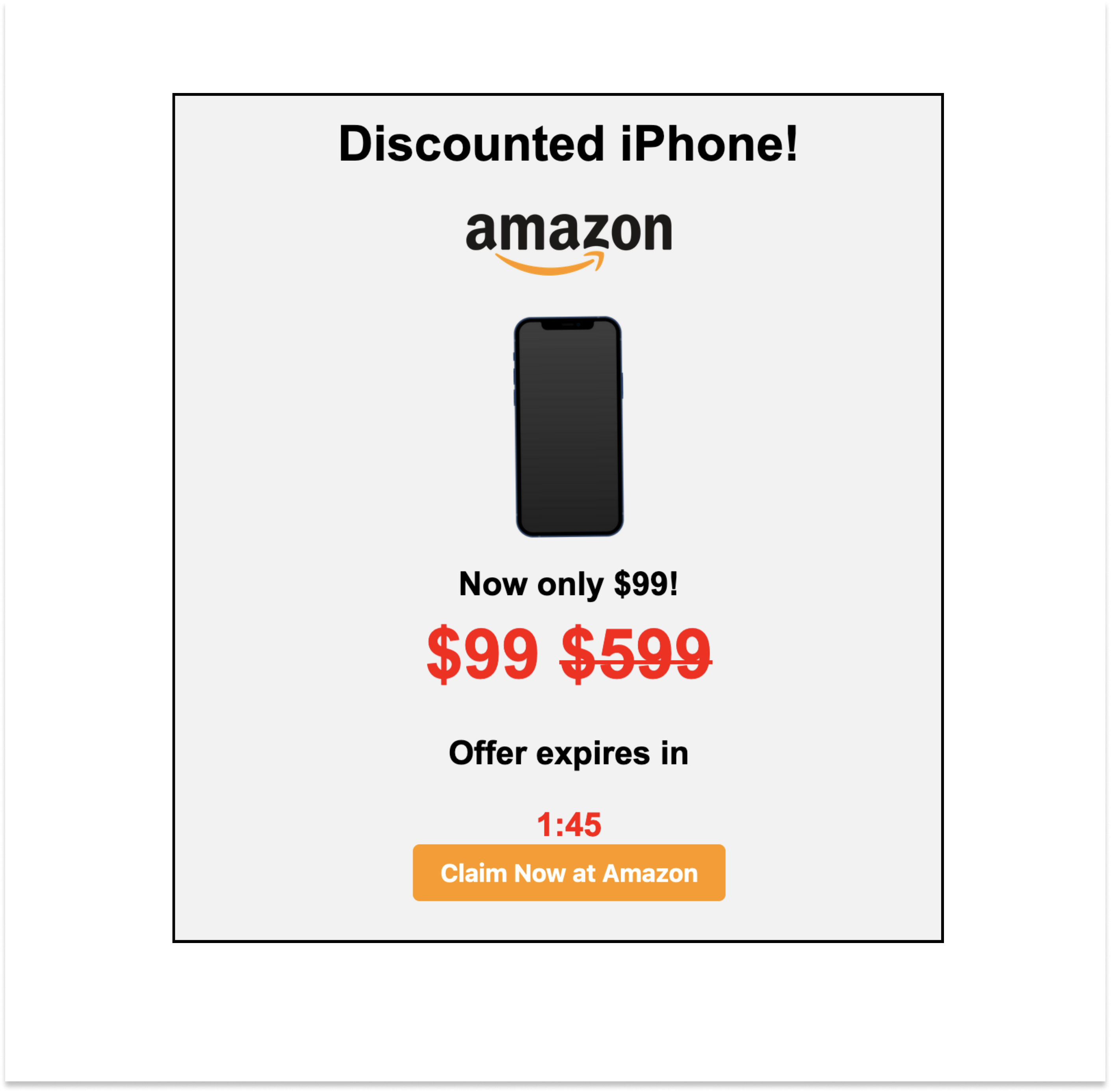}
\caption{Example of a clickjacking attack generated by ChatGPT which leads to an Amazon phishing page}
  \label{fig:clickjacking}
\end{figure}
The iFrame is an HTML tag that allows web developers to embed one webpage within another. Attackers use this feature to load a malicious website inside a legitimate one. These attacks are usually carried out in two ways: i) The iFrame is visible, and looks like it is part of the original website~\cite{secnhack-iframe-injection}, ii) The iFrame is hidden behind a legitimate object (such as a 'Submit' button) and the user is encouraged to interact with that button, wherein reality they are interacting with the hidden i-Frame. This technique is also known as \textit{clickjacking}~\cite{auth0-clickjacking}. While generating this attack, ChatGPT was also able to ensure that the iFrame object adhered to the same origin policy~\cite{portswigger-same-origin-policy} as to not trigger anti XSS (Cross Site Scripting) implementations in prevalent web browsers. Figure~\ref{fig:clickjacking} illustrates one such attack generated using ChatGPT which imitates an iPhone giveaway scam which has an hidden iFrame linking to a Amazon phishing page (Also generated by ChatGPT).

\subsection{Exploiting DOM classifiers}
\label{dom-classifiers}
Attackers can create phishing websites which are designed to avoid detection by a specific anti-phishing classifier. Liang et al.~\cite{liang2016cracking} identified and exploited several features of a website's Document Object Model (DOM), which are evaluated by Google phishing pages filter (GPPF), the client-side anti-phishing tool included with Google Chrome. A website's DOM is tree-like structure of objects created by a web browser when an HTML document is loaded, allowing programming languages like JavaScript to interact with and manipulate its elements. 
Liang et al.~\cite{liang2016cracking} found that modifying the parameters of these objects in a website can significantly reduce its chances of being detected by GPPF. \textit{ChatGPT} was able to create a phishing login page with all the modified parameters, which included changing input type attributes, removing relevant script elements, linking to target domains, sourcing external resources (such as images) from the root of the website, among many others.

\subsection{Polymorphic URL}
\label{poly-url}
\emph{Polymorphic phishing} attacks are designed such that a new URL is generated every time the website is accessed, making it hard for anti-phishing block-lists to evaluate and keep track of them. To accomplish this, attackers use PHP scripts executed on the server-side, that append a random string at the end of the URL each time there is a request to access the website~\cite{cofense-phishing-attack,lam2009counteracting}. Upon  appending this functionality to the template of our regular imitation attack (Section~\ref{attack_method}), ChatGPT was able to generate both the necessary client side HTML and server side PHP to emulate this attack. 

\subsection{Text encoding exploit} 
\label{punycode}
\begin{figure}[t]
\centering
\subfloat[HTML snippet generated by ChatGPT which imitates Yahoo! and encodes all visible text (including the title and form fields]{\includegraphics[width=0.4\textwidth]{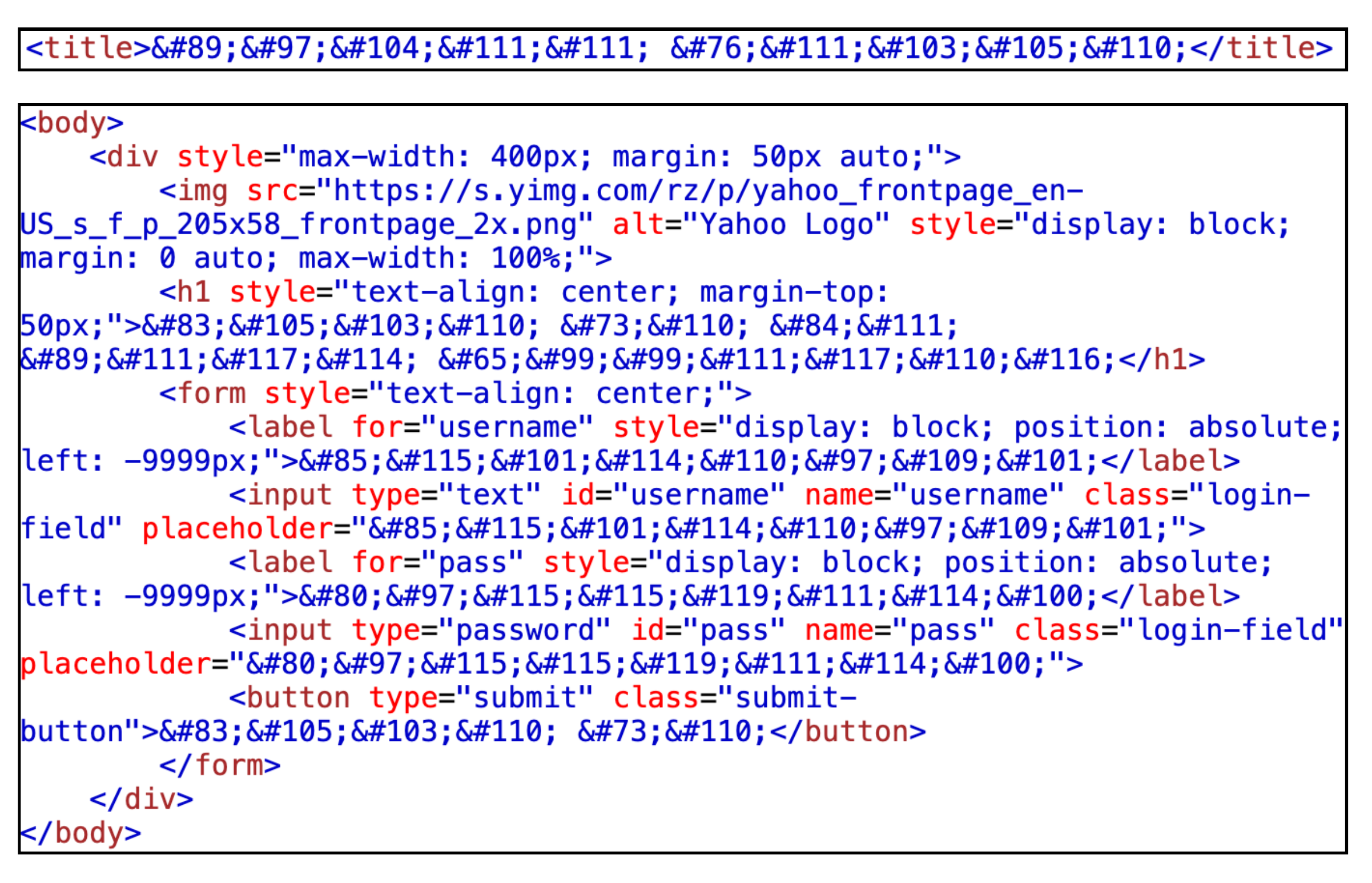}
\label{fig:punycode1}}\
\subfloat[The web-browser decodes the ASCII characters to display regular text]
{\includegraphics[width=0.28\textwidth]{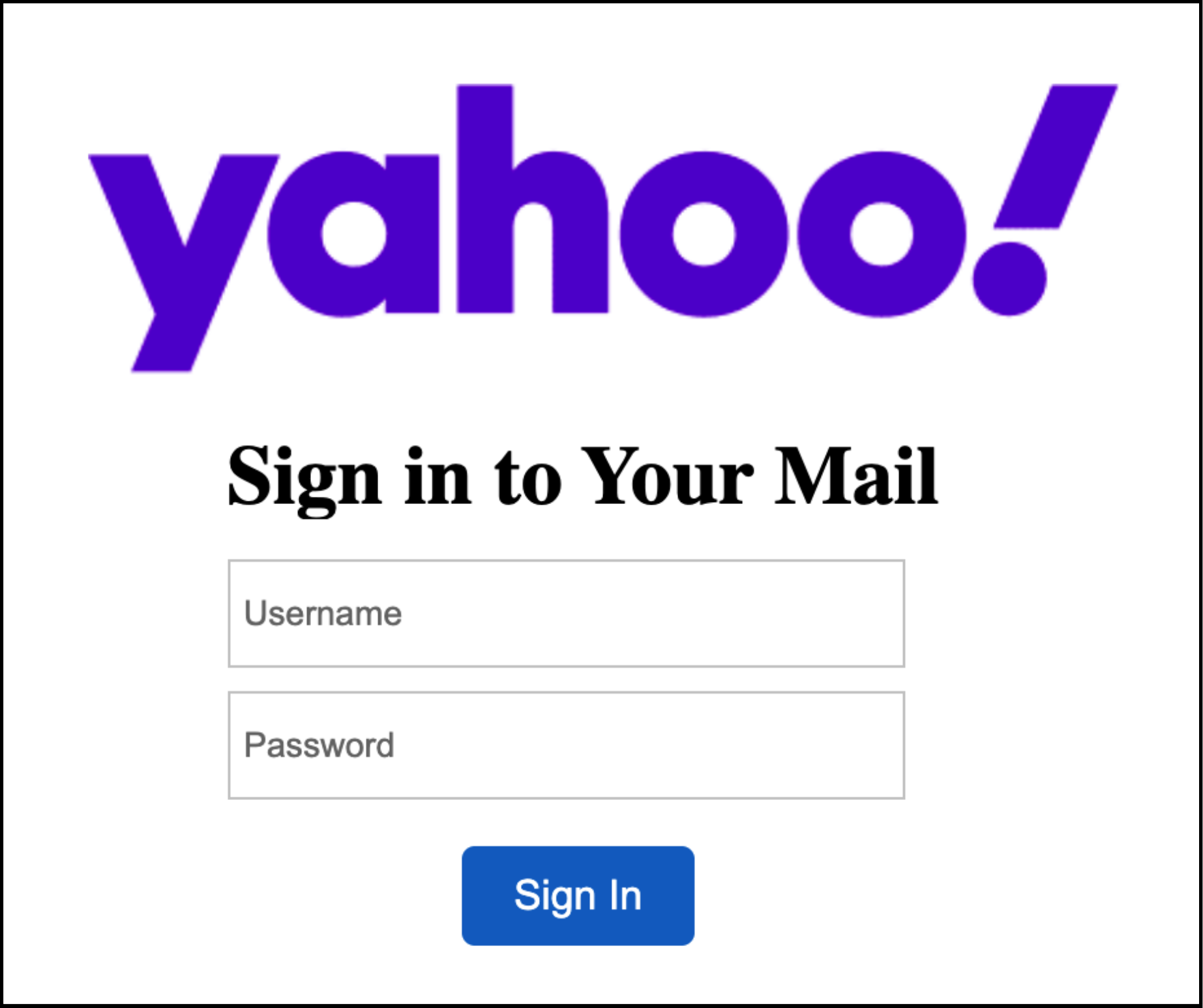}
\label{fig:punycode2}}
\caption{Generating a text encoding exploit using ChatGPT}
\label{fig:punycode}
\end{figure}
Attackers encode the text of the credential taking intention fields with ASCII or Morse code characters codes such that they are not recognizable from the website's source code~\cite{cybersecurityventures_punycode_phishing,fouss2019punyvis}. However, this code upon being rendered by a web browser, displays the text as intended. 
We were able to generate this by providing specific instructions on which fields to obfuscate in a way that it would be hard for text detection algorithms to detect. Figure~\ref{punycode} shows an example of ChatGPT generating a phishing website where the 'User-ID' and 'Password' fields are obfuscated in the source code. Figure~\ref{fig:punycode} illustrates one such attack created by ChatGPT.





\subsection{Browser fingerprinting attacks}
\label{browser-fingerprinting}
\begin{figure}[!ht]
\centering
  \includegraphics[width=0.8\columnwidth]{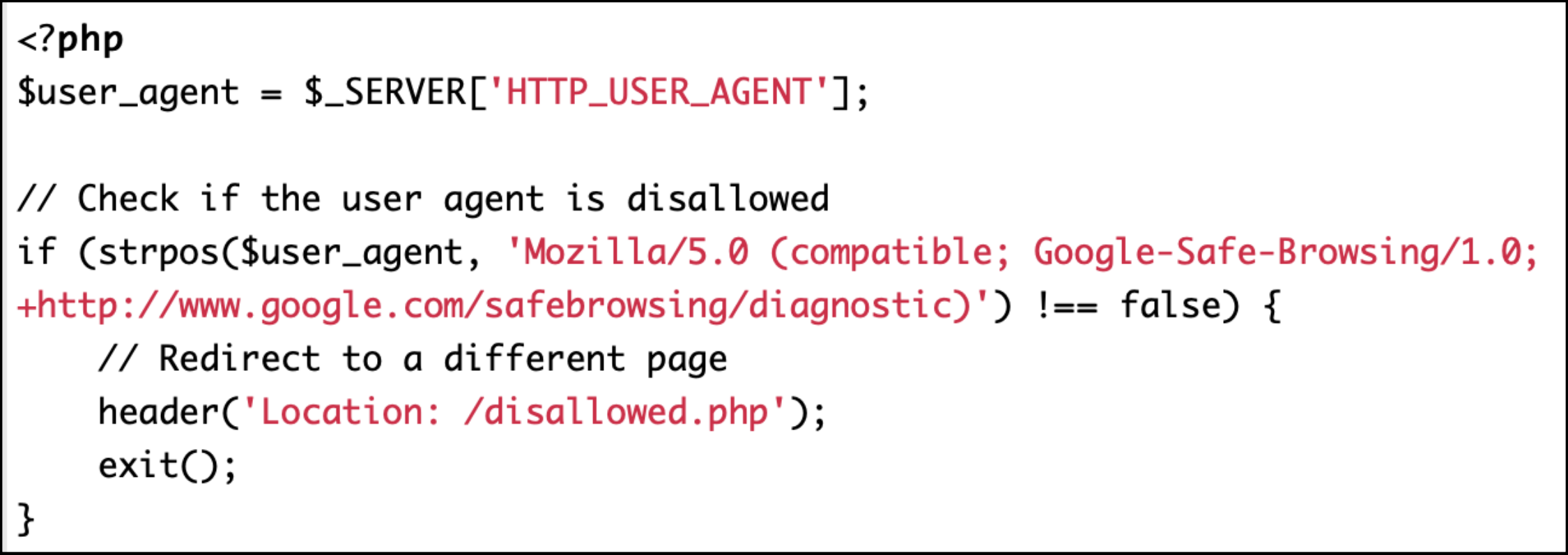}
\caption{Snippet of Server-side script generated by ChatGPT to evade crawling by Google Safe Browsing}
  \label{fig:browser-fingerprinting}
\end{figure}

A Browser Fingerprinting attack is designed such that the phishing attack is only shown to the user if they visit the website through a specific ``agent,'' i.e., web-browsers, mobile devices, or specific IP ranges~\cite{laperdrix2020browser,upathilake2015classification}.  Visiting such websites using other agents results in a benign or blank page being rendered. Such attacks are often used to evade anti-phishing bots which visit new websites to identify malicious features. Browser fingerprinting phishing attacks are becoming increasingly common and sophisticated, and can be difficult to detect and prevent. We were able to create prompts which could generate phishing pages which are rendered only when visited through a certain desktop web browser such as Google Chrome, or when it was accessed through a mobile device. Similarly we were able to restrict access to agents coming from certain IP ranges. Attackers can easily provide IP ranges belong to anti-phishing crawlers this way. Figure~\ref{fig:browser-fingerprinting} illustrates one such attack created by ChatGPT which can evade crawling by Google Safe Browsing~\cite{safebrowsing}.

\section{Iterative prompts and hosting the phishing attacks}
\label{iterative_hosting}
\begin{table}[]
\resizebox{0.95\columnwidth}{!}{
\centering
\begin{tabular}{c|ccc}
\hline
\multirow{2}{*}{\textbf{Attacks}} & \multicolumn{3}{c}{\textbf{Iterations}} \\ \cline{2-4} 
 & \multicolumn{1}{c|}{\textbf{Coder 1}} & \multicolumn{1}{c|}{\textbf{Coder 2}} & \textbf{Coder 3} \\ \hline
Design \& Credential transfer & \multicolumn{1}{c|}{11} & \multicolumn{1}{c|}{8} & 10 \\ \hline
Regular & \multicolumn{1}{c|}{+0} & \multicolumn{1}{c|}{+0} & +0 \\ \hline
Captcha phishing & \multicolumn{1}{c|}{+3} & \multicolumn{1}{c|}{+2} & +2 \\ \hline
QR Code phishing & \multicolumn{1}{c|}{+2} & \multicolumn{1}{c|}{+1} & +3 \\ \hline
Browser fingerprinting & \multicolumn{1}{c|}{+1} & \multicolumn{1}{c|}{+1} & +2 \\ \hline
DOM Features & \multicolumn{1}{c|}{+3} & \multicolumn{1}{c|}{+3} & +4 \\ \hline
Clickjacking & \multicolumn{1}{c|}{+5} & \multicolumn{1}{c|}{+3} & +4 \\ \hline
Browser-in-the-Browser & \multicolumn{1}{c|}{+5`} & \multicolumn{1}{c|}{+5} & 7 \\ \hline
Punycode & \multicolumn{1}{c|}{+1} & \multicolumn{1}{c|}{+1} & +3 \\ \hline
Polymorphic URLs & \multicolumn{1}{c|}{+1} & \multicolumn{1}{c|}{+2} & +4 \\ \hline
\end{tabular}}
\caption{Iterations required by our three coders to generate the phishing attacks using ChatGPT which imitate the Amazon brand. The first row signifies the number of iterations required to create the base design and transfer the credentials back to the attacker which is the same for all attacks, with the remaining rows signifying the number of iterations required on top of it to create the respective exploit.}
\label{iterations}
\end{table}

We examined the number of iterative prompts required by three independent coders (graduate students in Computer Science) to create each phishing attack discussed from Sections~\ref{regular-phishing-attacks} through ~\ref{browser-fingerprinting}. These coders possessed varying levels of technical proficiency in Computer Security: Coder 1 specialized in the field, Coder 2 had some familiarity through academic coursework, and Coder 3 had no prior exposure. Table~\ref{iterations} presents our findings.
Each coder created their own set of prompts for designing the website layout and for transmitting the stolen credentials back to the attacker, which they reused for multiple attacks.  
Our analysis reveals that all coders, regardless of their expertise in Computer Security, demonstrated similar performance when generating exploit prompts. This observation may suggest that crafting phishing attacks using ChatGPT does not necessitate extensive security knowledge, although it is important to note that all coders were technically proficient.
To further assess the ease with which ChatGPT can be used to generate phishing websites, it would be intriguing to expand this study to include participants who lack technical proficiency. This could provide a more comprehensive understanding of the potential risks associated with using ChatGPT for malicious purposes. 

We also deployed each attack on a free web hosting platform (000webhost~\cite{000webhost}) to evaluate the effectiveness of phishing website scripts produced through ChatGPT.  
The initial step involved incorporating the ChatGPT-generated scripts into their corresponding file formats (.html, .css, .js, etc.), followed by integrating these scripts into the default WordPress template supplied by the hosting service. To prevent unintended user access, we did not promote or sharing these websites in any way. Furthermore, we ensured that any data input into these sites was discarded, though no such instances were observed. Finally, we removed the websites soon after hosting them, again so that they cannot be accidentally accessed. 
\shirin{also remove them immediately after being tested?}\sayak{Yes}
Given their affordability, cybercriminals frequently leverage free hosting domains to execute large-scale phishing attacks~\cite{securityboulevard2021}. Utilizing ChatGPT-generated source code streamlines this process, further reducing the effort and technical expertise needed for successful deployment. 

%% file: conclusion.tex
\section{Discussion}
\label{discussions} 
Our work shows that ChatGPT can be used to generate both regular and evasive phishing attacks. While we focus on eight such attacks which are known to evade evaluation by anti-phishing entities, our approach can easily be extrapolated to other evasive attack categories. We discuss ramifications for the phishing ecosystem and limitations below: 
\newline
\newline
\textbf{Attacker opportunity:} Firstly, the process of generating and hosting phishing websites can be massively expedited using the ChatGPT API. As illustrated in Table~\ref{iterations}, very few prompts are required to generate the attacks, and the low API usage cost~\cite{openai-pricing} combined with the presence of free hosting domains~\cite{roy2022large}, makes the entire process very easy and cost-effective for the attacker. On the other hand, security vendors have, over time, designed extensive heuristic approaches to detect automated phishing kits. Presently, scrutiny of malicious content generated by Large Language Models such as ChatGPT is still at a nascent stage~\cite{si2022so}, thus making it more convenient for attackers to abuse this resource to release new attacks at a rapid pace. 
\newline
\shirin{let's remove this idea and keep it to ourselves. ;)}\sayak{Done}
\newline
\textbf{Missing resources:} We found that in some of the cases, external images included in the generated website might be outdated since the training cut-off date for ChatGPT was in October 2021~\cite{techcrunch-chatgpt-shrugged}. A follow-up prompt to replace these objects usually solved the problem. It can be assumed that attackers can also manually replace these resources whenever required or use ChatGPT implementations that have access to real-time internet resources such as AutoGPT~\cite{autogpt} or Bing Chat~\cite{autogpt}. 
\shirin{the following point was mentioned in the Measuring effectiveness. Let's merge them or remove one of them.}\sayak{Removed} 